# The effect of twitter-mediated activities on learning outcome and student engagement: A case study

Smaragdi Loutou, Nikolaos Tselios, Panagiota Altanopoulou
smaroloutou@gmail.com, nitse@ece.upatras.gr, galtanopoulou@gmail.com

Department of Educational Sciences and Early Childhood Education,
University of Patras, 26500, Rio, Patras

**Abstract**. The study reported in this paper, examines the effects of a rigorously designed introduction of Twitter in the educational process. In specific, it examines the relation between the students' use of Twitter in the context of well-organized educational activities and their grades in an academic course. The results demonstrated a significant correlation between the use of Twitter by the students and their performance in the course. Moreover, it emerged that their participation on Twitter, could improve the students' academic engagement and their sense of belonging to the academic community. Finally, possible effects of Twitter's use on students' perceived internet self-efficacy were also examined. The findings and their implications of the study are discussed in detail.

## Introduction

Social networks such as Facebook, and Twitter are a part of the citizens' daily life in the modern societies. According to Junco (2014a, p. 6), social networks are considered as services that let the user create and share data with other users. Social networks replaced extensively message sending and phone calls, especially for students. Digital devices are being used especially from youngsters for the use of the social networks which help them communicate with each other almost instantly and with low cost.

Moreover, educational use of Web 2.0 technology in general and social networks in particular could support effectively the process of learning and skills development (Altanopoulou, Tselios, Katsanos, Georgoutsou, & Panagiotaki, 2015, Altanopoulou, & Tselios, 2015). The application of Web 2.0 technology in education introduces significant opportunities due to their open nature, the facility and the support of students' cooperation and engagement (Hung & Yuen, 2010; Tselios, Altanopoulou & Komis, 2011, Katrimpouza, Tselios, & Kasimati, 2018). Alexander (2006) shows that Web 2.0 technology could effectively connect students with teachers and the learning materials. Therefore, it is suggested that Web 2.0 technology could support student-centered education involving students according to their skills and their personal interests. Also, as far as wiki technology is concerned, he believes that they are useful tools that help students to improve demanding competencies.

There are four advantages at the introduction of Web 2.0 in education and those are literacy, research, collaboration and the ability of publication (Crook, 2008). In specific, students have the ability for more personal research and organization of information. Those technologies offer tools that could enable students to learn in a personal way. Also, the users communicate with tools which fit at the education and improve collaboration. Finally, the technologies provide the ability to improve literacy in a way that students can express themselves and show their way of thinking. Digital tools provide new representation and expression possibilities with the use of different media like pictures, sound and video.

Twitter is one of those Web 2.0 technologies and it belongs in the general category of micro-blogging, which allows users to share posts by using a limited number of characters (140 and recently 280). An important characteristic of Twitter is the so-called “hashtag”. Hashtags are represented by the symbol of # and they are words or phrases without spaces. Users are able to follow a theme through hashtag where they could comment, share photos and videos etc. Twitter is a useful tool in real-time communication and the participation in it is influenced by peers (Costa, Bcham, Reinhardt & Sillaots, 2008). Twitter improves active learning and provides incentives to the students (Gao, Luo & Zhang, 2012). Students seem to gain confidence, reflection and collaborative learning and they improve their sociability (Chen & Chen, 2012; Smith & Tirumala, 2012).

In a study conducted by Tiernan (2013) involving 75 first-year students at the Dublin City University three basic research questions were examined, for the educational use of Twitter. Firstly, they examined if students adopted Twitter and on which motives. Secondly, the influence of Twitter uses on student engagement with the course. Finally, the students’ preferences regarding the different possibilities to integrate Twitter in the learning process.

Regarding the first question, students used twitter at a low rate due to technical difficulties and the absence of motivation. However, there was a high degree of use from shy and introvert students. The main reason was the feeling of comfort communication with a large number of users, without any fear or shame. As far as the second question is concerned, students reported that Twitter positively influences the interaction between the stakeholders of a course in four distinct ways. Firstly, they are able to submit questions for the course. Secondly, the brainstorming sessions between students. Finally, the third and the fourth way were about the comments and the conversations related to the courses from the students. For the third question, students mentioned three possible uses of Twitter in higher education. More specifically, 38% of the students want to use the tool not only in, but outside the classroom. 28% want to use Twitter as the official digital space of the course and 31% of students want to use it supplementary, as a complementary way to communicate with the educator.

Junco, Heibergert and Loken (2010) investigated the Twitter use by students and its influence on their academic engagement and their final grades. 125 students participated in a carefully designed study. The results demonstrated that Twitter can influence positively, both students’ engagement as well as their final grades.

Prestridge (2013) used Twitter to support freshmen and analyze types of interaction that arise, such as connectivity, academic culture and ingenuity. She identified a link between the student and the instructor. The most representative example of this link is the tweets by a student and the responds which occur by the instructor. Also, a novel academic culture emerged mainly by describing the content of the lectures by the students supported by publication of images, links and retweeting actions. The ingenuity, refers to the interaction of the student with the user interface of Twitter. The students while using this technology, experienced problems in applying academic protocols, in the limitation of 140 characters per tweet and while in the effort to handle multiple streams of tweets.

Moreover, Junco, Elavsky and Heiberger (2013) studied the systematic introduction of Twitter in the learning process. Researchers from the survey have used questions from “National Survey of Student Engagement” (NSSE) questionnaire, which assesses the students’ perceptions about their involvement in the academic community. In this way, they examined student participation prior the Twitter use in the classroom. Alongside Twitter, some of the students comprised a control group which used the tool Ning.  Junco, Elavsky and Heiberger (2013) used only 19 of NSSE questionnaire questions, which focus on the involvement of students in academic activities related to the curriculum and belong to the first of the five categories of NSSE. Questions comprising the NSSE questionnaire, collect information on five key issues, which are: (1) Participation in various academic activities, (2) Institutional requirements and nature of the courses (3) Perceptions of the University environment, (4) Assessment for educational and personal development from the beginning of the academic year (5) Demographics. The 19 questions used in the study are characterized by a significant degree of internal validity (Cronbach's alpha 0.80, Junco, Elavsky & Heiberger, 2013).

Subsequently, teachers asked students to read a particular book and provide a summary of it. The work should be carried out via Twitter and each student was requeted to comment on the posts of the students. Furthermore, they urged them to formulate questions for the course, to note important dates or to mark various events organized by the university community. In the first study participation through the tools was compulsory for all but in the second study participation was optional.

The research results reported by Junco, Elavsky and Heiberger (2013) are very encouraging for the prospects of educational use of Twitter. In specific, the results of the first study demonstrated that the group of Twitter had a higher learning gain in comparison to the control group, which used the Ning tool. In the second study, in which participation was voluntary and less organized, the results are not particularly showed significant difference in student performance. In conclusion, Junco, Elavsky and Heiberger (2013) showed that the use of Twitter tool, benefiting the students to participate in academic and performance. Moreover, it seems that teachers should be involved with the tool themselves to enhance the results of the students. Finally, it was found in this study that Web 2.0 technologies, and particularly Twitter, are effective when used in a proper rigorously designed instructional intervention framework.

From the aforementioned literature review, and despite the encouraging results obtained, it seems that the study on the effect of using the Twitter tool on the performance of students is still at its infancy. Therefore, it is necessary to study in a systematic manner the educational use of the tool and its implications.

In this study, the effectiveness of the activity design framework proposed by Junco, Elavsky and Heiberger (2013) is investigated in the context of an academic course. In specific, the purpose of this study was to investigate the effect of Twitter's use in the learning process. The research questions were:

RQ1. Is there a correlation between the students' Twitter use with their learning outcome (namely their final exam and final laboratory grade)?

RQ2. Is there a variation in students' engagement in the learning process?

In order to further investigate the first research question, the following secondary research questions made:

RQ1.1 Is there a correlation between students' performance with the help that students perceive?

RQ1.2 Is there a correlation between students' performance with their reported desire to incorporate Twitter in other subjects?

RQ1.3 Is there a correlation between students' performance with their overall Twitter activity in the context learning process?

RQ 1.4 Is there a correlation between students' self-efficacy and Internet use in the final performance in the laboratory?

RQ 1.5 Is there a correlation between students' performance in the laboratory, and how understandable they found the posed questions?

The rest of the paper is organized as follows: First the method of the study is presented followed by the analysis and presentation of the data obtained, organized per research question. Finally, the results of the study are discussed, and future research goals are introduced.

## Method of the study

### *Research method*

A one-group pretest–posttest design was adopted. The study took place from 12 May 2014 to 6 June 2016. The pretest comprised 43 questions. In specific, it included five demographic questions, 19 questions related to self-efficacy of Internet use and 19 questions from the NSSE instrument, translated in Greek. Similarly, the posttest questionnaire contained 46 questions, in which were added the questions related to the students' evaluation of their learning experience while the demographic questions were eliminated.

To measure Internet self-efficacy, a questionnaire was given which contained 19 questions in order to explore the students' improvement, if any, on their internet self-efficacy after their involvement with the Twitter (Hsu & Chiu, 2004; Papastergiou, 2010). The questions concerned the confidence felt by the students to navigate the Internet, the ability to use some Web 1.0 and Web 2.0 services and the ability to understand fundamental technical issues of Internet use (Hsu & Chiu, 2004; Papastergiou, 2010). The questions presented to the students were translated into Greek. The answers to the questions of self-efficacy instrument have had a 5-point Likert-like scale, ranging from 1 (strongly disagree) to 5 (strongly agree).

### *Participants*

The participants were university students attending a non-compulsory academic course, entitled "Introduction to Web Science". This particular course was offered in the second semester, in the Department of Educational Sciences and Early Childhood Education at the University of Patras. The research involved 17 students, who had chosen the abovementioned course. Moreover, the course instructor and the course lab assistants participated in the survey. The students were aged 18-34 years (mean = 21.76, SD = 3.63) with an average GPA score 7.5/10. From the participants, only one student had a Twitter account at the beginning of the course.

### *Materials research*

The online questionnaire service, SurveyMonkey (www.surveymonkey.com) was adopted to create and distribute the study questionnaires and Google Docs for the announcement of the activities. The software packages which had been used for the analysis were Excel 2010, SPSS

v21.0 as well as the Wordle tool, which is a tool for visual presentation of textual data. Finally, the Topsy[1] tool was used, which allows review of students' publications in the relevant hashtag of course, regardless of the time that has passed, allowing quite sophisticated queries and comparisons.

*Procedure and Activity Description*

As for the involvement of students on Twitter, they should respond to questions which were related to each session of the laboratory, using the hashtag (#websci). In detail, the questions per week were:

*1st week*: "Please indicate the pros and cons of the heuristic evaluation (a method to evaluate the usability of a web site)".

*2nd week*: "Why do we use the time sequence of keystrokes to a proxy and not count it with the law of Fitts?"

3rd week: "Which is the most important function of Moodle? Also, identify the positives and negatives of Moodle in relation to a Wiki".

4th week: "Which is the main advantage of virtual worlds compared to other online games? Moreover, mention a positive and a negative element for the adoption of Second Life in the educational process".

Every week, the participants had at their disposal 24 hours to answer each question. Throughout their involvement, students had to follow three rules. Firstly, to respond to the question that had given to them with no more than two tweets. Secondly, to comment to the replies of at least two fellow students each week and thirdly to submit at least two questions related to the course during the semester.

Since participation was optional, an incentive was given by the course instructor to students in order to participate. According to Hoinville and Jowell (1978), the incentives need special attention and it should be used as "a further gesture." It shouldn't be seen as a reward, but as a gift gesture for participation. In this activity, the incentive was one additional point in the final grade of the course for the student with the highest activity. Moreover, the next two students with the most intense participation received half point in the final grade of the course. These students were informed for their achievement at the end of the activity process.

## Results

*Analysis of student's involvement on Twitter*

The participation of students is examined by the type of tweets that made and their total number of comments per week. Regarding the type of activities, students were making tweets in order to comment the activities, to provide questions or to discuss with other students.

Table 1 presents the student participation on the total number of tweets they made. In specific, students began the first week with the highest participation rate of 27.76% of the total responses. Over the next two weeks there was a gradual decrease, while in the fourth week there was a small increase, where students perform to ¼ of the total responses for four weeks. Table 1 provides the results for each type of students' tweets.

[1] www.topsy.com

**Table 1.** Twitter actions per week and per type of comments

| **Tweets** | **1st Week** | **2nd Week** | **3d Week** | **4th Week** | **Tweets of activities** | **Tweets of comments** | **Tweets of questions** |
|---|---|---|---|---|---|---|---|
| **Total** | 181 | 158 | 150 | 163 | 132 | 484 | 37 |
| **Average** | 10.65 | 9.29 | 8.82 | 9.24 | 7.76 | 28.4 | 2.18 |
| **SD** | 10.73 | 12.64 | 9.32 | 10.21 | 4.82 | 35.6 | 2.20 |
| | **27,76%** | **24,24%** | **23%** | **25%** | **20,21%** | **74,1%** | **5,62%** |

An interesting note is the fact that the largest average in comments (mean = 28.47, SD = 35.61), concerns the category "tweets for comments". Students showed particular preference in making comments and debating with other students for the activities, which constitute 74.1% of the total comments. Instead, the comments relating to questions were 5.62% of the total comments.

Additionally, the number of tweets for each category counted beyond the minimum required participation. In the category "tweets for activities" students made only two tweets. In the category "tweets for comments" 59% (10/17) of students made 348 additional tweets beyond the minimum requirement. Even after the end of each activity the category "tweets for comments" gather a greater involvement of students, showing a tendency to discuss with their peers. Finally, in the category «tweets for questions" 29% of students made additional tweets beyond the minimum requirement (5 to 17 students).

For the analysis of the type of students' tweets a tool of lexicography used namely Wordle. Below the results presented by the students' comments for each of the micro-activities. Figures relating to the comments for each week respectively.

The 1$^{st}$ week, students discussed the laboratory for the method of "Heuristic Evaluation" and were asked to report a positive and a negative element. Figure 1 shows the meanings given by the students. The responses appear to be associated with those given in the literature concerning the method of heuristic evaluation. Specifically, the main aim of heuristic evaluation is to reduce the number of criteria (Nielsen & Molich, 1990), which is assigned by the responses of students to the concept of "minimalist". Still, the main advantage of the method according to the Holleran (1991) is inexpensive to implement, which is discussed by students as "low cost". As far as disadvantages of the method are concerned, the Avouris (2000) refers to the subjective nature of the method, which is based on empirical rules and findings, which indicated the students by "subjective" regarding the negative method. Therefore, students seem to respond, which is in agreement with the international literature.

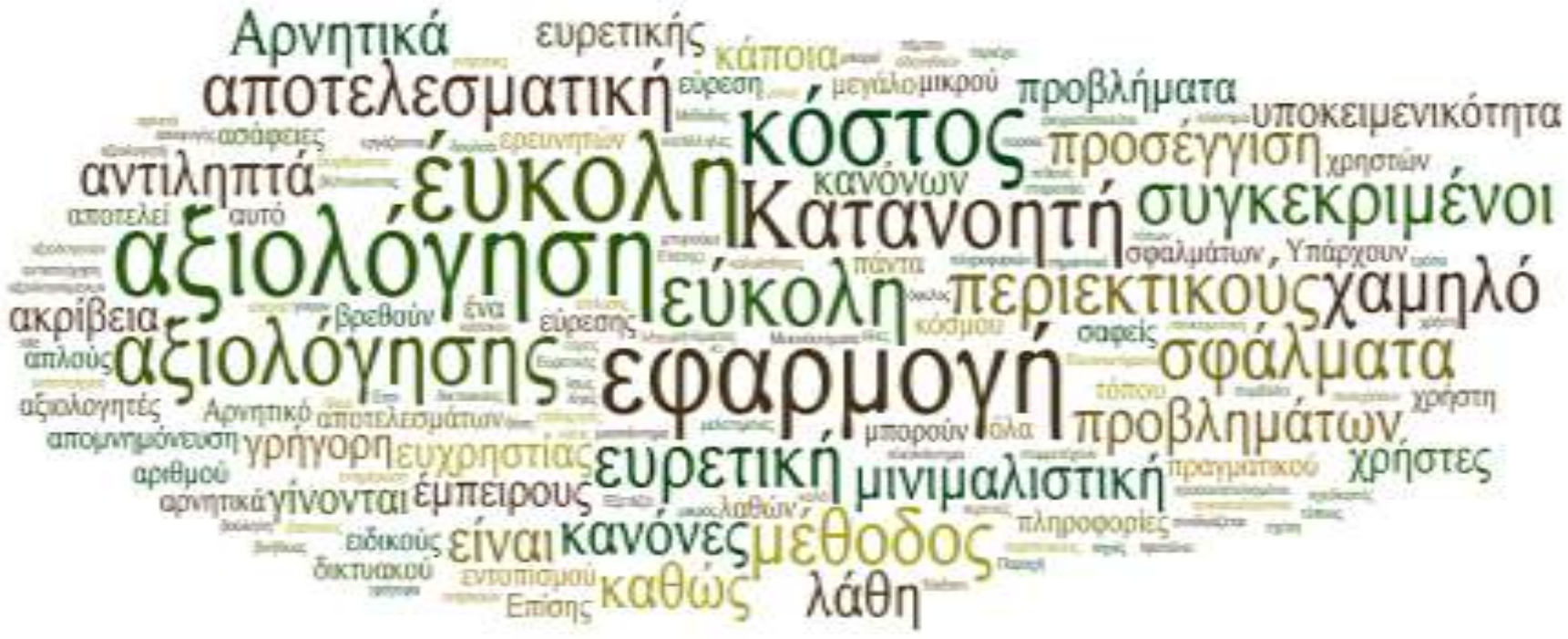

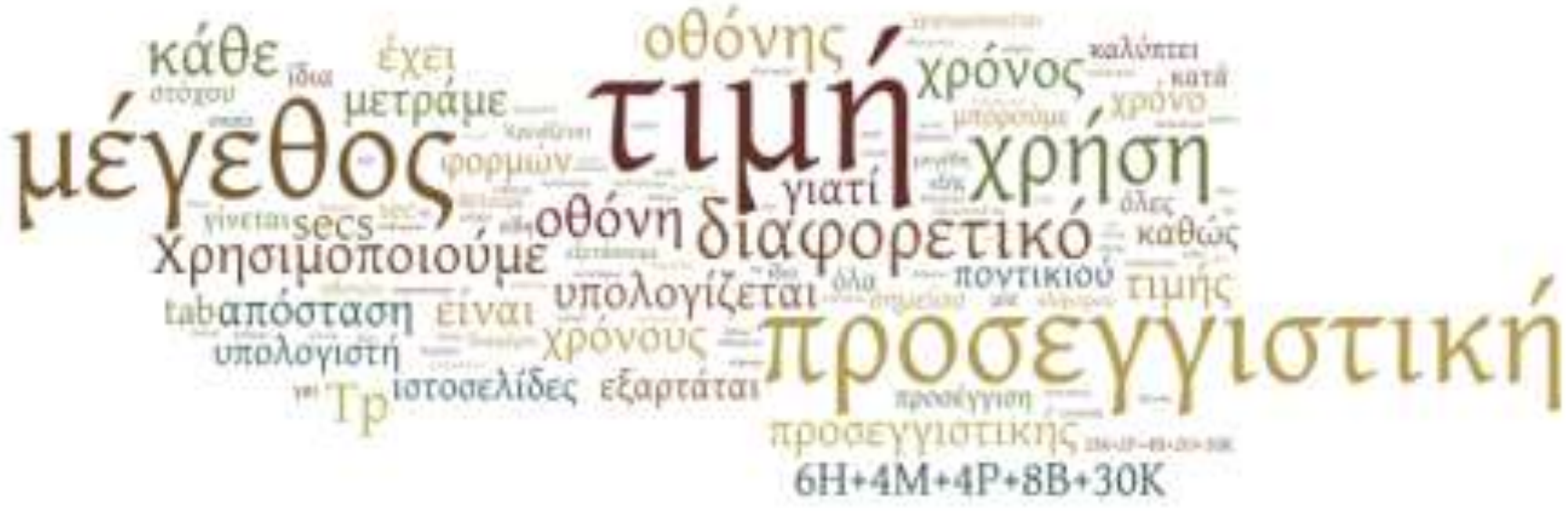

**Figure 1.** Results of lexicography for the 1st activity and the 2nd activity

In the second week (Figure 1) students were asked to comment on why an approximate value is used in order to calculate the time (Tp) and to calculate the time needed to complete a form. At this point it is interesting to study the first part of the question, which concerns the opinion of the students to use an approximate value. Students referred that approximate value is useful because "the size of each screen is different". This is in line with other surveys that said that the time varies, depending on the distance of the target, in which the cursor is to be directed and in the dimensions of the target (Avouris, 2000).

In the third week, students were asked to indicate which Moodle’s service is important and the advantages and disadvantages compared with the Wiki technology. It was found (Figure 2), that Moodle tool is a type of "diary" and a tool to complete "activities." Indeed, Moodle is characterized for the flexibility offered in doing activities (Cole & Foster, 2007). Other important services that Moodle has are tools for synchronous and asynchronous communication, questionnaires and quizzes (Rice, 2011) and less in the "diary", which reviewed the students. Moreover, students believed that Moodle is integrated in 'distance education' which is confirmed by other researchers (Dobrzański, Honysz & Brytan, 2006; Bayne & Ross, 2014).

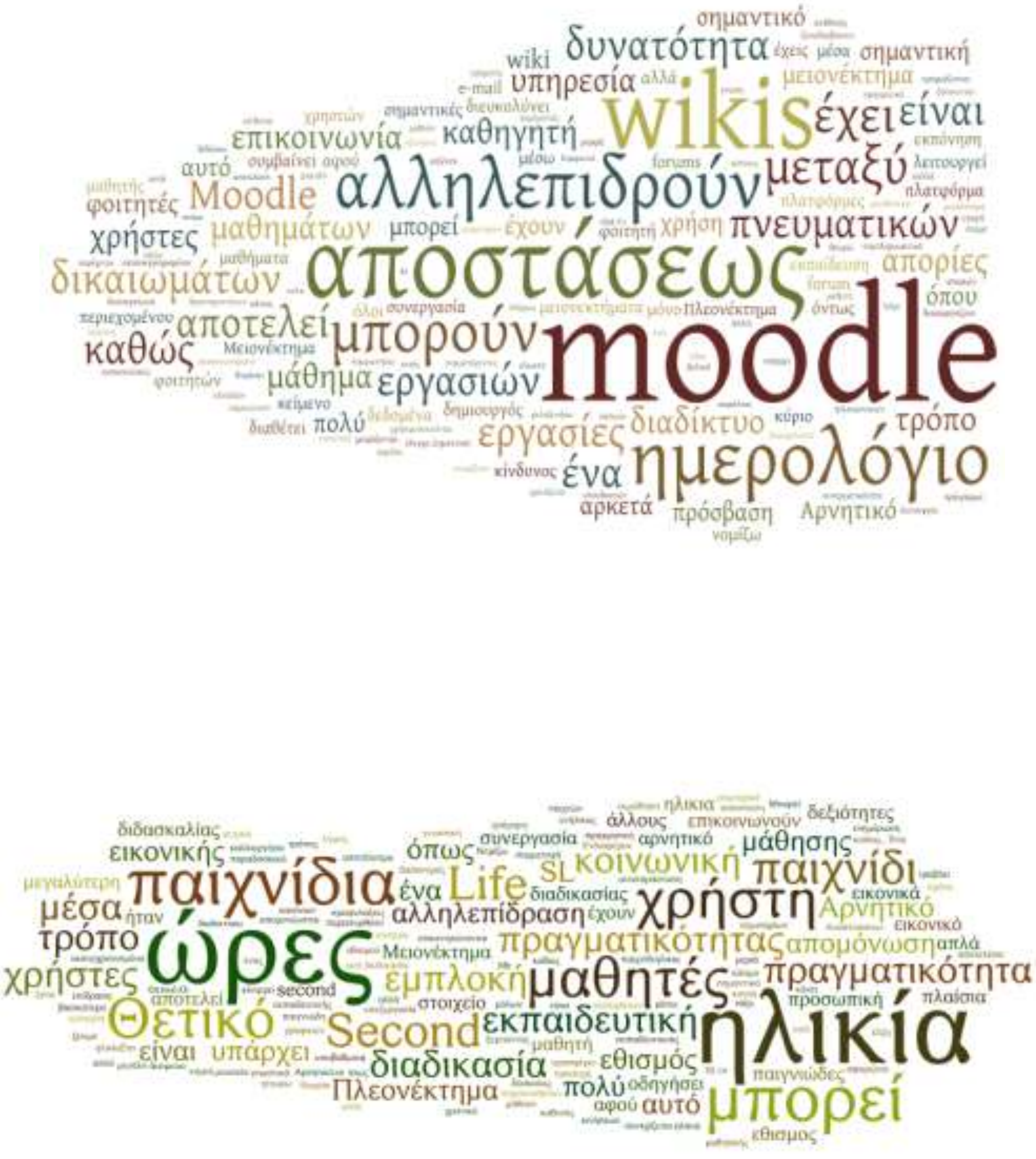

**Figure 2**. Results of lexicography for the 3d and the 4th activity

In the last week, participants were asked to report the main advantage of virtual reality games compared to other online games. Then, they had to indicate a positive and a negative element in the use of Second Life, at the educational process. Students' comments (Figure 2) were related strongly with two words: age and hours. Also, as shown in Figure 2, students often used the words: isolation, addiction, learning, gaming, interaction and playful. Students' answers which related with high interaction is confirmed by other surveys (Kallonis & Sampson, 2011). Also, Second Life or other three-dimensional systems (e.g. Sloodle), seems to increase students' engagement in learning process with students having an active role in learning (Kallonis & Sampson, 2011).

## *Twitter use and learning outcome*

The first research question of the study, (correlation between the use of the tool of the course and the students' learning outcome) it has been examined in two dimensions. The first dimension refers to the the final exam test grade and the second to the final grade of the laboratory, which in turn is the average grade of four compulsory mini-projects presented to the students. The students received a mean laboratory grade of 8.85 (SD = 1,35), a mean final exam grade of 6.17 (SD = 1.6) and a mean final grade of 7.44 (SD = 1.01). The final degree is a weighted average of laboratory grade (40%) and final exam grade (60%). \

As to the degree of final examination appear to be highly correlated with the total comments on microactivity (Pearson's r = 0,66, p = 0,00 <0,05, s). For the final grade of the laboratory appear to be highly correlated with the total number of comments made by the students (Pearson's r = 0,63, p = 0,00 <0,05, s). Still, the queries wording set via Twitter, seems to have a high correlation with the grade of the final written test (Pearson's r = 0,56, p = 0,02 <0,05, s). Finally, the total activity on Twitter seems to confirm the above, since it shows a high correlation with the degree of the final exams (Pearson's r = 0,59, p = 0,01 <0,05, s). For a more complete investigation of the first question, raised and secondary research questions as mentioned previously. The first of these relates to the correlation between the overall academic average of the students and the assistance said it offered them the Twitter tool to the subject. Significant correlation (Pearson's r = 0,58, p = 0,01 <0,05, s). Also, there seems to be a high correlation of the overall academic average of students with their desire for integration of the tool in other subjects and is statistically significant (Pearson's r = 0,52, p = 0,00 <0,05, s).

Yet another research question concerns the extent of the overall academic average student and the possible relation to the total activity in the Twitter. It appears that the most good students involved more in the activities of Twitter (Pearson's r = 0,57, p = 0,01 <0,05, s).x

The next research question concerns the possible association between diversification in self-efficacy use of the Internet and the ultimate performance in the lab. Interestingly, there is a negative correlation between variables (Pearson's r = -0,48, p = 0,05, s). It seems that students already high skills in using the Internet, and thus little variation, are the students who were efficient in class within the laboratory part. Conversely, students with greater diversification in self-efficacy are the students who were less efficient in the laboratory.

The last research question relates to the relationship between the final grade in the laboratory and how understandable were the questions to the students. It found a high correlation between the two variables (Pearson's r = 0,62, p = 0,00 <0,05, s).

### *Difference on the levels of students' academic engagement*

In the second research question, "if there is variation in NSSE questionnaire after student engagement on Twitter of course," there seems to be significant differentiation. In the initial test, the average was 2.11 and the standard deviation of 0.29, while in the final test, the average score was 2.26, and the standard deviation 0.26. Therefore, there is a positive differentiation at NSSE questionnaire and was statistically significant (paired-sample T-test, p = 0,02 <0,05, s).

**Table 2.** Total of comments at Twitter per week and per type of comments

| **Differentiation in students' engagement** | **Pre-test Score** | **Post-test Score** |
|---|---|---|
| Faculty members | 2,06 | 2,53 |
| Administrative staff | 1,94 | 2,06 |
| Students | 2,71 | 2,88 |
| **Totally** | **2,11** | **2,26** |

Interestingly, the study of the variations with respect to each class of members of the academic community. Such differentiation was examined in relations between students with faculty members, administrative staff and other students. Table 2 presents these differences on average, which concern the relations of the students, with each category separately.

The fact that the teaching of the course participated in the activities, led the students to make the greatest improvement, which amounts to 22.8%. Then, in relations with administrative staff, there is a little differentiation, by 0.12 points (mean pre-test score = 1.94, mean post-test score = 2.06), although they did not participate in any of the activities. Finally, on their relations with the other students seem to differentiated by 6.27%.

## Conclusions

In this study the integration of Twitter in the educational process was investigated. In specific, Twitter has been used in the laboratory section of a course in a Department of Education at the University of Patras for four weeks. The students had to follow specific rules set by the researchers regarding the Twitter use. The results suggest that students, who participated to every week's activities on Twitter, were more efficient in class compared to those who did not participate (both in the laboratory and in the final exams). This is in agreement with the results reported by Junco, Elavsky and Heiberger (2013) who showed that the inclusion of Twitter in the educational process led to an increase to the students' grade. On the other hand, they showed that students who used Twitter in an optional base did not appear to have any increase to their grades.

An important finding is that the students' involvement in Twitter, can lead to a variation in their self-reported academic engagement, even in a small time frame of four weeks. Similar results were presented by Elavsky and Heiberger (2013) who showed that there is an increase in students' involvement in academic community, when using Twitter compulsory and with encouragement, but over a longer period. Also, they showed that the group who used optionally the tool, had no difference in the NSSE questionnaire. It is noteworthy that students improved student their relationships with the teachers.

As for the students' self-efficacy, it was found a moderate correlation with the overall participation of students. Also, it was found a moderate correlation with the statement that students are better organized their thinking due to the limitation of characters that Twitter required.

In addition, it was investigated any relation between the students' grade point average and their involvement in Twitter. It was found a high correlation of the grade point average with answers regarding the self-assessment process. Specifically, the higher average they have, the more want to introduce Twitter in other subjects. Still, there is a high correlation with how understandable were the questions in the process of activities and how they considered that benefited in the subject, their involvement in Twitter. An important finding is the fact that the students of higher performance made the most tweets. From these findings it appears that higher level students are more open and positive to new ways of learning.

Future work could build upon the findings of this study. In specific, an extension to conduct studies with a broader profile of participants is proposed with diversification of subjects. It is also proposed to conduct longitudinal studies to closer monitor students' possible variations in their attitudes towards the perceived usefulness of Twitter. In the present study, it is expected that there is no significant variation in students' perception for their ability to use the Internet. However, the use of the tool for a longer period of time is likely to significantly improve the students' self-efficacy. In addition, the investigation of learners' behavioral intention to adopt well organized educational activities mediated by social networks using technology acceptance models constitutes an additional research goal (Altanopoulou & Tselios, 2017, Tselios, Daskalakis, & Papadopoulou, 2011, Daskalakis, & Tselios, 2011). Finally, it would be interesting if a more systematic and deeper analysis of the type of students' interventions was conducted by using suitable Educational Data mining and Learning Analytics techniques (Kotsiantis, Tselios, Filippidi & Komis, 2013).